\documentclass[preprint,12pt]{elsarticle}

\usepackage{amssymb}
\usepackage{amsmath}

\usepackage{booktabs}
\usepackage{bm,bbm}
\usepackage{xcolor}
\usepackage{hyperref}
\usepackage{graphicx}

\newcommand{\blue}[1]{\textcolor{blue}{#1}}

\newcommand\pkg[1]{\texttt{#1}}
\let\proglang=\textsf 

\graphicspath{{./}{../graphics/}}

\journal{}

\newcommand{\red}[1]{\textcolor{black}{#1}}

\begin{document}
	
	\begin{frontmatter}
		
		
		
		\title{CALF-SBM: A Covariate-Assisted Latent Factor 
		Stochastic 
			Block Model} 
		
		
		\author[label1]{Sydney Louit} 
		\author[label2]{Evan A.\ Clark}
		\author[label2]{Alexander H.\ Gelbard}
		\author[label3]{Niketna Vivek}
		\author[label1]{Jun Yan}
		\author[label4]{Panpan Zhang}
		
		\affiliation[label1]{organization={Department of Statistics, 
				University of Connecticut},
			addressline={215 Glenbrook Rd, U-4120}, 
			city={Storrs},
			postcode={06269}, 
			state={CT},
			country={USA}}
		
		\affiliation[label2]{organization={Department of 
		Otolaryngology, 
				Vanderbilt University Medical Center},
			addressline={1211 Medical Center Dr}, 
			city={Nashville},
			postcode={37232}, 
			state={TN},
			country={USA}}
		
		\affiliation[label3]{organization={School of Medicine, 
		Vanderbilt 
				University},
			addressline={1161 21st Ave S}, 
			city={Nashville},
			postcode={37232}, 
			state={TN},
			country={USA}}
		
		\affiliation[label4]{organization={Department of 
		Biostatistics, 
				Vanderbilt University Medical Center},
			addressline={2525 West End Ave}, 
			city={Nashville},
			postcode={37203}, 
			state={TN},
			country={USA}}

		\begin{abstract}
			We propose a novel network generative model extended
			from the standard stochastic block model by concurrently
			utilizing observed  node-level information and accounting 
			for
			network-enabled nodal heterogeneity. The proposed model 
			is so 
			called covariate-assisted latent factor stochastic block 
			model 
			(CALF-SBM). The inference for the proposed model is done 
			in a 
			fully Bayesian framework. The primary application of 
			CALF-SBM in 
			the present research is focused on community detection, 
			where a 
			model-selection-based approach is employed to estimate 
			the 
			number of communities which is practically assumed 
			unknown. To 
			assess the performance of CALF-SBM, an extensive 
			simulation 
			study is carried out, including comparisons with multiple 
			classical and modern network clustering algorithms. 
			Lastly, the 
			paper presents two real data applications, respectively 
			based on 
			an extremely new network data demonstrating collaborative 
			relationships of otolaryngologists in the United
			States and a traditional aviation network data containing
			information about direct flights between airports in the 
			United
			States and Canada.
		\end{abstract}
		
		
		
		
	\end{frontmatter}
	
		
		
		\section{Introduction}
		\label{sec:intro}

		Community detection garners widespread interest in network 
		data
		analysis, with important applications in various fields such 
		as 
		social
		networks~\citep{handcock2007model, ouyang2023amixed}, brain
		networks~\citep{sporns2016modular}, input-output
		networks~\citep{wang2021regional}, citation
		networks~\citep{vaneck2014citnetexplorer}, and recommendation
		systems~\citep{liu2024variational}. Broadly, there are two
		predominant classes of approaches to network community 
		detection (or
		network clustering). The first class involves probabilistic
		graphical models characterizing network community structure,
		including stochastic block models~\citep{zhao2017asurvey,
			abbe2018community}, latent space 
			models~\citep{hoff2002latent,
			handcock2007model}, and random dot product
		graphs~\citep{athreya2017statistical}, among others. The 
		second
		class defines a network-metric-based objective function for 
		each
		possible community structure, and then finds the community
		structure that optimizes the objective function. Most 
		commonly used
		metrics are modularity~\citep{newman2006modularity} and its
		variants~\citep{ouyang2020clique, shang2020anovel}.
		
		Most community detection methods focus primarily on
		network topology, represented by the adjacency
		matrix, but often overlook node-level attributes or edge
		features that could be informative. A pertinent question 
		arises: can
		community detection accuracy improve by integrating auxiliary
		information from nodes or edges? Several recent studies
		addressed this by combining network topology with node
		attributes. \citet{yang2013community} developed a scalable
		algorithm for binary-valued node attributes, later adapted for
		multilayer networks by \citet{contisciani2020community}.
		Concurrently, \citet{zhang2016community} introduced a joint 
		community detection criterion that balances node connections 
		and 
		features through a tuning parameter. More recently, 
		\citet{yan2021covariate} targeted sparse networks using a 
		covariate 
		regularized community detection algorithm, a concept 
		subsequently 
		extended to multilayer networks by \citet{xu2022covariate}. 
		\red{Other recent developments in community detection 
		algorithms 
			accounting for covariates include the following: 
			\citet{binkiewicz2017covariate} improved spectral 
			clustering by 
			leveraging node covariates; \citet{zhang2019node} 
			proposed a 
			stochastic block model based on the fusion of block-wise 
			effects and 
			node features; \citet{relvas2023amodelbased} used node 
			covariates to 
			model cluster centroids; \citet{hu2024network} considered 
			a 
			community detection method using node covariates modified 
			by network 
			connectivity.} To the best of our knowledge, research 
			simultaneously 
		considering node covariate information and nodal 
		heterogeneity for 
		network generation process under the general stochastic block 
		modeling framework is scarce.

		In this paper, we propose a novel stochastic block model that
		leverages observed node-level covariates influencing network
		connectivity and latent node-level variables determining nodal
		heterogeneity. The model aims to enhance
		clustering accuracy by incorporating auxiliary node 
		information.
		The inference is done in a Bayesian framework. The number of
		communities is assumed unknown, and will
		be chosen through a standard but effective model selection
		procedure. Our model is distinct from a very recent
		algorithm~\citep{shen2023bayesian}, which is based on a
		covariate-dependent random partition model without taking 
		nodal
		heterogeneity into account. Our main contribution
		is a new network generative model accounting for both
		node-level covariates and nodal heterogeneity.
		Through comparisons with popular competing methods and
		real data applications, we illustrate the importance of 
		taking these
		factors into account in network clustering. Our 
		implementation is
		publicly available via a user-friendly \proglang{R} package
		\proglang{calfsbm} on GitHub.

		The rest of the manuscript is organized as follows:
		Section~\ref{sec:model} introduces a novel stochastic block 
		model
		simultaneously accounting for node-level covariates and
		heterogeneity. Section~\ref{sec:est} provides the details of a
		Bayesian estimation
		algorithm for the proposed model, including prior distribution
		derivations, a practical strategy for hyperparameter 
		initialization,
		a post-sampling procedure necessary for model identifiability 
		and an
		effective approach for estimating community number when it is
		unknown. Simulations are presented in Section~\ref{sec:sim}, 
		which
		is split into two subsections respectively for known and 
		unknown
		community numbers. Two real network data applications based 
		on an
		otolaryngologist collaboration network and an airport 
		reachability
		network are discussed in Section~\ref{sec:app}, followed by
		concluding remarks and discussions about the limitations and 
		future
		works in Section~\ref{sec:dis}.

		\section{Model}
		\label{sec:model}

		Given an undirected network $G := (V, E)$ with node set~$V$ 
		and edge
		set~$E$, let $n = |V|$ be the number of nodes in $G$. By 
		convention,
		the structure of $G$ is characterized by the associated 
		adjacency
		matrix $\bm{A}:=(A_{ij})_{n \times n}$ with $A_{ij} = 1$ 
		indicating
		the existence of an edge between nodes $i \in V$ and $j \in 
		V$;
		$A_{ij} = 0$, otherwise. Self-loops are not considered 
		throughout.
		For each $i \in V$, we define a membership vector $\bm{Z}_i :=
		\{Z_{i1}, Z_{i2}, \ldots, Z_{iK}\} \in \{0, 1\}^{K}$, where 
		$Z_{ik}
		= 1$ indicates that node~$i$ belongs to community $k \in [K] 
		= \{1,
		2, \ldots, K\}$. The number of communities~$K$ is unknown. The
		vector~$\bm{Z}_i$ is typically modeled by a multinomial 
		distribution:
		$\bm{Z}_i \sim \text{Multinomial}(1, \bm{\alpha}_i)$, where
		$\bm{\alpha}_i := (\alpha_{i1}, \alpha_{i2}, \ldots, 
		\alpha_{iK})$
		is a hyperparameter vector with $\alpha_{ik} \in [0, 1]$ and
		$\sum_{k = 1}^{K} \alpha_{ik} = 1$. The membership indicator 
		for
		node~$i$ (denoted as~$\pi_i$) is fully dependent on 
		$\bm{Z}_i$,
		defined as $\pi_i:= \arg \max_{k \in [K]} \bm{Z}_i$ for each 
		$i$. In
		a standard stochastic block
		model~\citep[SBM,][]{nowicki2001estimation, 
		abbe2018community}, the
		probability of the existence of an edge between nodes~$i$ 
		and~$j$
		follows a Bernoulli distribution with probability 
		$\eta_{\pi_i,
			\pi_j} \in [0, 1]$, which depends on the communities that 
			$i$ and
		$j$ belong to. Given $\pi_i = k$ and $\pi_j = l$ with $k, l 
		\in
		[K]$, we write $\eta_{\pi_i, \pi_j}$ as $\eta_{i, j \mid k, 
		l}$ for
		simplicity and clarity.

		In the present research, the standard SBM is extended from two
		aspects. First, we incorporate observed node-level 
		information into
		the model. For each $i \in V$, we consider a $p$-dimensional
		covariate vector~$\bm{X}_i$, which contains auxiliary 
		information
		that may contribute to node connectivity pattern. For any 
		pair of
		nodes $i, j \in V$, we define a dual measure $S_{ij} := 
		g(\bm{X}_i,
		\bm{X}_j) \in \mathbb{R}$ to quantify the similarity between 
		the two
		nodes based on their covariates, where, for the special case 
		of 
		$\bm{X}_i = \bm{X}_j$, the function $g(\bm{X}_i, \bm{X}_j)$ 
		yields 
		$0$. Moreover, the function $g(\cdot)$ is symmetric for 
		undirected 
		networks. \red{Then we model the probability $\eta_{i, j 
				\mid k, l}$ using a $\text{logit}$ function as 
				follows:}
		\[
		\text{logit}(\eta_{i, j \mid k, l}) = \beta_0 + \beta_{kl} 
		S_{ij},
		\]
		where $\beta_0$ presents a baseline commonality across all
		communities, and $\beta_{kl}$ specifies the homogeneity of 
		edge
		connection preserved between communities~$k$ and~$l$. For
		compactness, let $\bm{\beta}$ denote the collection of 
		$\beta_0$ and
		all $\beta_{kl}$'s.

		The second extension involves incorporating node-level latent
		factors that reflect nodal heterogeneity and their propensity 
		for
		connections. Let $\theta_{i}$ denote the latent factor for 
		node $i
		\in V$, presenting its tendency of generating links. Then,
		$\eta_{i, j \mid k, l}$ is characterized by
		\begin{equation}
			\label{eq:eta}
			\text{logit}(\eta_{i, j \mid k, l}) = \beta_0 + 
			\beta_{kl} S_{ij}
			+ \theta_{i} + \theta_{j},
		\end{equation}
		where large, positive values of $\theta_{i}$ and $\theta_{j}$
		jointly promote the connection between~$i$ and~$j$. For
		tractability, we assume that $\theta_{i}$'s, for $i = 1, 2, 
		\ldots, 
		n$,
		are independent and follow a normal distribution with
		mean $0$ and unknown variance $\sigma^2$ \red{as suggested 
			by~\citet{hoff2003random}.}

		To summarize, we propose a covariate-assisted latent factor
		stochastic block model (CALF-SBM) that simultaneously 
		accounts for
		observed node-level characteristics and latent node-level
		factors. Given the number of communities $K$, 
		Equation~\eqref{eq:eta}
		defines the rate parameter $\eta_{i, j \mid k, l}$ for a 
		Bernoulli
		trial characterizing the presence of an edge between nodes~$i$
		and~$j$. That is
		\begin{equation}
			\label{eq:chsbm}
			\Pr(A_{ij} = 1 \mid \pi_i = k, \pi_j = l, \bm{\beta},
			\theta_{i}, \theta_{j}, \sigma^2 ; K, \bm{X}_{i}, 
			\bm{X}_j)
			= \eta_{i, j \mid k, l}.
		\end{equation}

		In addition to the parameters in
		$\bm{\Theta} := (\bm{\beta}^\top, \bm{\theta}^\top, 
		\sigma^2)^\top$,
		we are interested in $\bm{Z}$ which
		determines community structure and $K$ which is unknown in 
		practice.
		The inference of $\bm{Z}$ is done together with the other 
		parameters
		in $\bm{\Theta}$ through a Bayesian estimation framework. To 
		infer
		$K$,
		we employ a model selection procedure. Both will be detailed 
		in a
		subsequent section. Consequently, the observed data 
		likelihood,
		assuming that $K$ is known, is formulated under the standard
		assumption of conditional independence given membership 
		information.
		The likelihood of the proposed CALF-SBM with a given~$K$ is
		\begin{equation}
			\label{eq:full_lik}
			\mathcal{L}(\bm{\Theta}, \bm{Z} \mid \bm{A}; K, \bm{X}) =
			\prod_{i <
				j} \eta_{i, j \mid k, l}^{A_{ij}}(1 -\eta_{i, j \mid 
				k, 
				l})^{1 -
				A_{ij}},
		\end{equation}
		where $\eta_{i, j \mid k, l}$ is defined in 
		Equation~\eqref{eq:eta}.
		Likelihood estimation of Equation~\eqref{eq:full_lik} is
		computationally challenging due to the large number of 
		parameters
		and the marginal likelihood derivation of each parameter (or 
		latent
		factor) requires a high dimensional integration. Therefore, we
		adopt a Bayesian estimation approach. We would like to note 
		that the
		proposed model can be extended to directed networks with 
		effortless
		modifications.

		\section{Bayesian Estimation}
		\label{sec:est}

		We consider a fully Bayesian estimation for the proposed 
		model using
		a Markov Chain Monte Carlo (MCMC) algorithm, for which $K$ is 
		fixed, 
		followed by a model selection procedure to determine the 
		optimal $K$ 
		value.

		\subsection{Prior and Posterior}
		\label{sec:prior_posterior}

		In our model, we impose hierarchical prior structures in 
		order to
		make the prior assumptions vague assuming known $K$. Inspired 
		by~\citet{handcock2007model}, we propose the following prior 
		distributions to complete the Bayesian specification:
		\begin{enumerate}
			\item[(1)] $\bm{\beta} := (\beta_0, \beta_{11}, 
			\beta_{12}, 
			\ldots,
			\beta_{KK})\sim \text{MVN}(\bm{\mu},\bm{\Sigma})$,
			\item[(2)] $\sigma^2 \sim \text{IG}(a, b)$,
			\item[(3)] $\bm{\alpha}_{i} := (\alpha_{i1}, \alpha_{i2}, 
			\ldots,
			\alpha_{iK}) \overset{\text{i.i.d.}}{\sim} 
			\text{Dir}(\bm{\gamma})$,
		\end{enumerate}
		where $\text{MVN}(\bm{\mu},\bm{\Sigma})$ represents the 
		multivariate
		($((K^2 + K)/2 + 1)$-variate in our model) normal 
		distribution with
		mean vector $\bm{\mu}$ and covariance structure $\bm{\Sigma}$,
		$\text{IG}(a, b)$ represents the Inverse-Gamma distribution 
		with
		shape parameter $a$ and scale parameter $b$, and
		$\text{Dir}(\bm{\gamma})$ represents the Dirichlet 
		distribution with
		concentration parameters $\bm{\gamma}$. The symbols 
		$\bm{\mu}$,
		$\bm{\Sigma}$,
		$a$, $b$, and $\bm{\gamma} := (\gamma_1, \gamma_2, \ldots, 
		\gamma_K)$
		comprise our hyperparameters. All hyperparameters are 
		specified by
		the users based on actual available prior knowledge about 
		unknown
		parameters and latent factors.

		The MCMC algorithm iterates over the model parameters and 
		latent
		factors with the given priors. For those with full conditional
		distributions in closed forms, we sample them directly via 
		Gibbs
		sampling; otherwise we adopt a Metropolis--Hastings
		algorithm~\citep{hastings1970mh}. For succinctness, we use 
		`others'
		to denote all the unknowns other than the parameter or latent 
		factor
		explicitly specified in the formulas. The full conditional 
		posterior
		distributions are given as follows:
		\begin{enumerate}
			\item[(1)] $p(\bm{\beta} \mid \text{others}) \propto
			\phi(\bm{\beta} ; \bm{\mu},\bm{\Sigma}) 
			\mathcal{L}(\bm{\Theta},
			\bm{Z} \mid \bm{A}; K, \bm{X})$;
			\item[(2)] $p(\theta_i \mid \text{others}) \propto
			\phi(\theta_i; 0,\sigma^2)
			\mathcal{L}(\bm{\Theta}, \bm{Z} \mid \bm{A}; K, \bm{X})$;
			\item[(3)] $\sigma^2 \mid \text{others} \sim \text{IG}(a 
			+ n/2,
			b + \sum_{i = 1}^{n} \theta_{i}^2 / 2)$;
			\item[(4)] $p(\bm{Z}_i \mid \text{others}) \propto
			f(\bm{Z}_i ; \bm{\alpha}_i)
			\mathcal{L}(\bm{\Theta},
			\bm{Z} \mid \bm{A}; K, \bm{X})$;
			\item[(5)] $\bm{\alpha}_i \mid \text{others} \sim
			\text{Dir}(\bm{\gamma} + \bm{Z}_i)$;
		\end{enumerate}
		where $\phi(\cdot)$ in (1) is the density of a
		$((K^2 + K)/2 + 1)$-variate normal distribution,
		$\phi(\cdot)$ in (2) is the density
		of a univariate normal distribution, and $f(\cdot)$ in (4) is 
		the
		density of a Dirichlet distribution, respectively.

		The prior distributions were set to be vague but proper to 
		ensure
		proper posterior distributions. In particular, for the 
		regression
		coefficients, we set $\bm{\mu} = \bm{0}$ and
		$\bm{\Sigma} = 100\bm{I}$, where $\bm{I}$ is the identity 
		matrix.
		This prior distribution covers a wide range of regression 
		coefficient
		values. Then
		for $\sigma^2$, we set $a = b = 1$ such that the associated
		Inverse-Gamma distribution has an infinite mean, and meanwhile
		provides a sufficiently large variation for $\theta_i$ values.
		Lastly, we sample each $\gamma_i$ in $\bm{\gamma}$ 
		independently
		from $\text{Beta}(1, K)$, which provides large but
		not excessive variability in cluster size.

		\subsection{Initialization Strategy}

		Appropriate initial values are critical for the convergence of
		MCMC-based algorithms. We propose an effective strategy for 
		initial
		value specifications to greatly reduce the number of 
		iterations and
		computing time for the proposed MCMC algorithm.

		Specifically, we set the initial node memberships (i.e,
		$\pi_i^{\text{(ini)}}$ for $i = 1, 2, \ldots, n$) by 
		implementing a
		$K$-medians clustering algorithm~\citep{pkg:kmedians} to the
		observed data. The initial value of $\theta_i$, for
		$i = 1, 2, \ldots, n$, depends on the observed degree of node 
		$i$ 
		(denoted $d_i$), given by $\theta_i^{\text{(ini)}} =
		\log((d_i / (\sum_{i = 1}^{n} d_i / n)) + c)$,
		where $c = 10^{-4}$ is an offset constant
		ensuring the validity of logarithm function (especially for
		isolated nodes with $d_i = 0$). Given 
		$\pi^{\text{(ini)}}_i$'s and
		$\theta^{\text{(ini)}}_i$'s, we ran a logistic
		regression~\citep{pkg:arm} on observed adjacency matrix 
		$A_{ij}$'s with calculated $S_{ij}$'s (based off fully 
		observed 
		$\bm{X}_i$'s and $\bm{X}_j$'s) and $(\theta^{\text{(ini)}}_i 
		+\theta^{\text{(ini)}}_j)$'s as covariates to get the initial 
		values 
		for the regression coefficients: $\bm{\beta}^{\text{(ini)}}$.

		\subsection{Implementation}
		
		The MCMC was implemented with \proglang{R} package
		\pkg{NIMBLE}~\citep{pkg:nimble}, which provides an interface 
		of
		programming algorithms using modeling languages from the BUGS
		(Bayesian inference Using Gibbs Sampling) project. 
		\pkg{NIMBLE} is
		particularly popular for generating samples from the 
		complicated
		posterior distributions that are not in closed forms via the
		Metropolis--Hastings algorithm. Meanwhile, \pkg{NIMBLE} is 
		appealing
		for automatically determining available conjugate families to
		improve sampling efficiency. For more detailed features and
		advantages of \pkg{NIMBLE}, we refer the interested readers
		to the pioneering paper by~\citet{devalpine2017programming}.

		The codes for practical implementation are made publicly 
		available in
		an \proglang{R} package, which collects all the
		functions needed to fit the CALF-SBM (including diagnostic 
		tools).  
		The package, called \pkg{calfsbm}, is available on GitHub:
		\url{https://github.com/sydneylouit539/calfsbm}.

		\subsection{Post-sampling Procedure}

		Like many other network models for community detection, the 
		proposed
		CALF-SBM experiences non-identifiability since the
		likelihood described in Equation~\eqref{eq:full_lik} is 
		invariant to
		the permutation of node membership labels $\bm{Z}$. This 
		particular
		dilemma is referred to as the label-switching
		problem~\citep{stephens2000labelswitching}. Thus, a 
		post-processing
		procedure must be implemented prior to drawing inference for
		$\bm{Z}$.

		Specifically, we post-process the posterior MCMC samples by 
		applying
		the artificial identifiability constraints method. More 
		precisely,
		we re-organize the rows and columns of the coefficient matrix
		$\bm{\beta}$ according to the constraint:
		$\beta_{11} < \beta_{22} < \cdots < \beta_{KK}$ for each 
		posterior
		sample,
		followed by a re-assignment of individual membership in 
		accordance
		with the re-ordered $\beta_{kl}$'s. In practice, the 
		implementation
		is done by using a built-in function from the
		\pkg{label.switching} package~\citep{pkg:labelswitching}, 
		which
		is available in the \proglang{CRAN}. We find that this 
		approach is
		particularly useful for smaller networks with fewer available 
		data.

		\subsection{Number of Clusters}
		\label{sec:estK}

		In the majority of existing research for community detection, 
		the
		number of clusters, $K$, is assumed to be known a
		priori~\citep{zhang2016community, ouyang2023amixed}. However, 
		this
		assumption appears unrealistic in practice. Estimating the 
		number of
		clusters itself is a challenging problem that has received a 
		great
		deal of attention.

		Recent research~\citep{geng2019probabilistic, 
		shen2023bayesian}
		proposed a dynamic framework that estimates $K$ in an 
		iterative
		manner, together with other unknown parameters and latent 
		factors.
		However, as mentioned by~\citet{geng2019probabilistic}, these
		methods tend to create small extraneous communities with 
		roughly
		even size, resulting in overestimating $K$, which is also 
		observed
		in our simulations in Section~\ref{sec:sim}. Besides, the 
		change of
		$K$ causes the change of dimension of regression coefficients 
		for
		the proposed model, giving rise to additional challenges for
		algorithm convergence (even with an informative prior on $K$).

		Thus, we resort to a model selection approach as performed
		by~\citet{handcock2007model}, where the fundamental idea is 
		to make
		inference about unknown parameters and latent factors with a 
		wide
		range of $K$ values, and then select the optimal value of $K$
		according to some information criterion. While the Bayesian
		information criterion (BIC) was selected
		by~\citet{handcock2007model}, we choose the widely applicable
		information 
		criterion~\citep[WAIC,][]{watanabe2010asymptotic}, since
		the WAIC averages over the posterior distribution instead of
		conditioning on a point 
		estimate~\citep{gelman2013understanding},
		making it more helpful for models with hierarchical and 
		mixture
		structures like the proposed CALF-SBM. The calculation of 
		WAIC can
		be conveniently done via a built-in function from 
		\pkg{NIMBLE}.

		\section{Simulation Studies}
		\label{sec:sim}

		In this section, we carry out extensive simulations to assess 
		the
		goodness-of-fit of the proposed model and its performance 
		against
		popular competing methods in the literature. The simulation 
		study is
		divided into two parts: known $K$ and unknown $K$. For each,
		different competing methods are considered alongside
		sensitivity analyses that will be elaborated explicitly. The
		\proglang{R} codes for all the simulations are available in 
		the
		online supplements with well-documented random numbers for
		reproducibility.

		\subsection{Data Generation}
		\label{sec:data_gen}

		Motivated from the two real network data in 
		Section~\ref{sec:app},
		we generated synthetic networks with $n \in \{200, 400\}$ and
		$K \in \{2, 3, 4\}$, with $100$ networks being generated for 
		each
		combination of $n$ and $K$. Concurrently, node membership 
		$\bm{Z}_i$,
		for $i = 1, 2, \ldots, n$, was drawn from an
		unbalanced multinomial distribution of size $K$ and 
		probability
		parameters proportional to $(K, K - 1, \ldots, 1)^{\top}$.
		The intercept term $\beta_0$ was fixed to $1$, leading to 
		network
		density of approximately $10\%$. We set the within-cluster 
		effects
		$\beta_{kk}$, for $ k = 1, 2, \ldots, K$ to equally-spaced 
		values
		between $-1.6$ and $-1$. Assuming $K = 4$, for instance, the 
		values
		of $\beta_{kk}$ for $k = 1, 2, 3$ and $4$ were respectively 
		set to
		$-1.6$, $-1.4$, $-1.2$ and $-1$. On the other hand, all the
		between-cluster effects (i.e., $\beta_{kl}$ with $k
		\neq l$) were fixed to $-3$, resulting in a smaller 
		probability of
		establishing links across nodes from different communities.

		Given the clustering configuration,  we generated the
		$p$-dimensional covariates $\bm{X}$ in two steps. First, we 
		chose
		$K$ equally spaced points on the surface of a $p$-dimensional 
		ball
		with radius $\sqrt{2\omega}$ ($\omega > 0$) as the centers of 
		$K$
		communities. Since these selected points were invariant in 
		rotation,
		we, in practice, proposed to use the polar coordinate system 
		for the
		center selection; For instance, the points denoted by
		$(\sqrt{2\omega}, 2k\pi / K)$ with $k = 1, 2, \ldots, K$ were 
		chosen
		for $p = 2$. Second, for $i = 1, 2, \ldots, n$, we 
		independently
		drew $\bm{X}_i$ from a
		$p$-variate normal distribution with mean at the selected 
		point
		corresponding to the clustering membership (of node $i$) in 
		the
		first step and covariance matrix $\bm{I}_p$. Since the
		variance-covariance structure of $\bm{X}$ is fixed, the ball 
		radius
		$\sqrt{2 \omega}$ determines the degree of cluster overlapping
		explained by $\bm{X}_i$'s from different communities, with a 
		larger
		value of $\omega$ signifying a more significant community 
		structure
		governed by observed covariates, and consequently indicating a
		stronger signal-to-noise ratio. For each pair of generated
		$\bm{X}_i$ and $\bm{X}_j$, we employed the Euclidean distance 
		to
		construct the covariate-based similarity structure $S_{ij}$.

		After generating latent factors presenting node connectivity
		tendency using $\theta_i \overset{\text{i.i.d.}}{\sim}
		\mathcal{N}(0, 0.3)$, we created $A_{ij} = A_{ji}$ based on 
		the
		proposed model given in Equation~\eqref{eq:chsbm} for each 
		pair of
		distinct nodes $i$ and $j$. Accordingly, the generated 
		networks were
		undirected. Throughout the simulation, we set $A_{ii} = 0$ 
		for all
		$i = 1, 2, \ldots, n$ to exclude self-loops. For each 
		replicate, the 
		adjacency matrix and node covariates are assumed to be 
		observed, 
		whereas the true clustering and nodal heterogeneity 
		information are 
		latent.

		\subsection{Results}
		\label{sec:res}

		In this section, we present our simulation results that help 
		assess
		the performance of the proposed CALF-SBM as well as its 
		comparison
		to other competing methods in the literature. For each study, 
		a
		total of \red{$200$} independent synthetic network data were 
		generated
		according to the approach detailed in 
		Section~\ref{sec:data_gen}.
		For each simulation run for CALF-SBM, we fixed the burn-in 
		number
		and the (follow-up) number of iterations to $5000$ and $10^4$,
		respectively. The thinning parameter was set to $10$ to 
		reduce the
		auto-correlation in the posterior samples. In addition, three 
		chains
		were run simultaneously for a combination of posterior 
		samples. We
		employed the Gelman--Rubin 
		diagnostic~\citep{gelman1992convergence}
		alongside trace plots to examine the convergence of posterior
		samples, where we found that the total number of iterations 
		under
		the current setting was sufficiently large to ensure 
		convergence.

		\subsubsection{Known $K$}
		\label{sec:knownK}

		The first task is to evaluate the performance of the 
		algorithm for
		recovering the parameters in the CALF-SBM based on the 
		\red{$200$}
		independent replicates of simulated networks. We ran 
		simulations for
		both $n = 400$ and $n = 200$ with fixed $\omega = 1.5$, and 
		observed
		similar results, so only present those for $n = 400$ with
		$K \in \{2, 3, 4\}$ in Table~\ref{tab:sim_undirected}. 
		Specifically, 
		we provide bias, standard error (SE), empirical standard 
		deviation
		(ESD) and coverage percentage (CP) for each scenario
		(e.g., $\{n = 400, K = 2\}$), and conclude that the estimates
		from the proposed algorithm are unbiased and consistent.

		\begin{table}[tbp]
			\label{tab:n_400_est}
			\centering
			\caption{Estimation summary for $n = 400$ with $K = \{2, 
			3, 4\}$
				and $\omega = 1.5$. SE: standard error; ESD: 
				empirical 
				standard deviation; CP: coverage percentage ($\%$).}
			\label{tab:sim_undirected}
			\footnotesize
			\setlength{\tabcolsep}{3.6pt}
			\begin{tabular}{c rrrr rrrr rrrr}
				\toprule
				& \multicolumn{4}{c}{$K = 2$} & \multicolumn{4}{c}{$K 
				= 3$}
				&
				\multicolumn{4}{c}{$K = 4$} \\
				\cmidrule(lr){2-5} \cmidrule(lr){6-9} 
				\cmidrule(lr){10-13}
				Para. & Bias & SE & ESD & CP & Bias & SE & ESD & CP & 
				Bias &
				SE & ESD & CP \\
				\midrule
				$\beta_0$ & 0.006 & 0.044 & 0.042 & 95 & 0.001 & 
				0.045 &
				0.051 & 93
				& 0.005 & 0.051 & 0.051 & 94 \\
				$\beta_{11}$ & $-0.003$ & 0.026 & 0.025 & 95 & 0.000 &
				0.038 		& 0.039 & 94 & $-0.025$ & 0.047 & 0.061 & 
				86
				\\
				$\beta_{12}$  & $-0.002$ & 0.041 & 0.039 & 95 & 
				$-0.001$ &
				0.059 & 0.065 & 92 & 0.000 & 0.086 & 0.138 & 97 \\
				$\beta_{22}$ & $-0.001$ & 0.026 & 0.023 & 97 & 
				$-0.004$ &
				0.031 & 0.032 & 93 & $-0.013$ & 0.049 & 0.048 & 96 \\
				$\beta_{13}$ & & & & & $-0.005$ & 0.059 & 0.049 & 98 &
				$-0.022$ & 0.086 & 0.073 & 98 \\
				$\beta_{23}$ & & & & & 0.001 & 0.058 & 0.062 & 95 &
				$-0.025$ & 0.089 & 0.133 & 96 \\
				$\beta_{33}$ & & & & & 0.001 & 0.024 & 0.021 & 96 &
				$-0.006$ & 0.045 & 0.043 & 94 \\
				$\beta_{14}$ & & & & & & & & & $-0.013$ & 0.087 & 
				0.075 &
				97 \\ $\beta_{24}$ & & & & & & & & & 0.004 & 0.092 & 
				0.140
				& 97 \\
				$\beta_{34}$ & & & & & & & & & $-0.002$ & 0.089 & 
				0.089 &
				96 \\
				$\beta_{44}$ & & & & & & & & & $0.005$ & 0.046 & 
				0.050 &
				93 \\
				$\sigma$ & $-0.001$ & 0.015 & 0.015 & 97 & 0.001 & 
				0.017 &
				0.017 & 97 & 0.003 & 0.012 & 0.010 & 98 \\
				\bottomrule
			\end{tabular}
		\end{table}

\begin{figure}[tbp]
			\centering
			\includegraphics[width=0.9\textwidth]{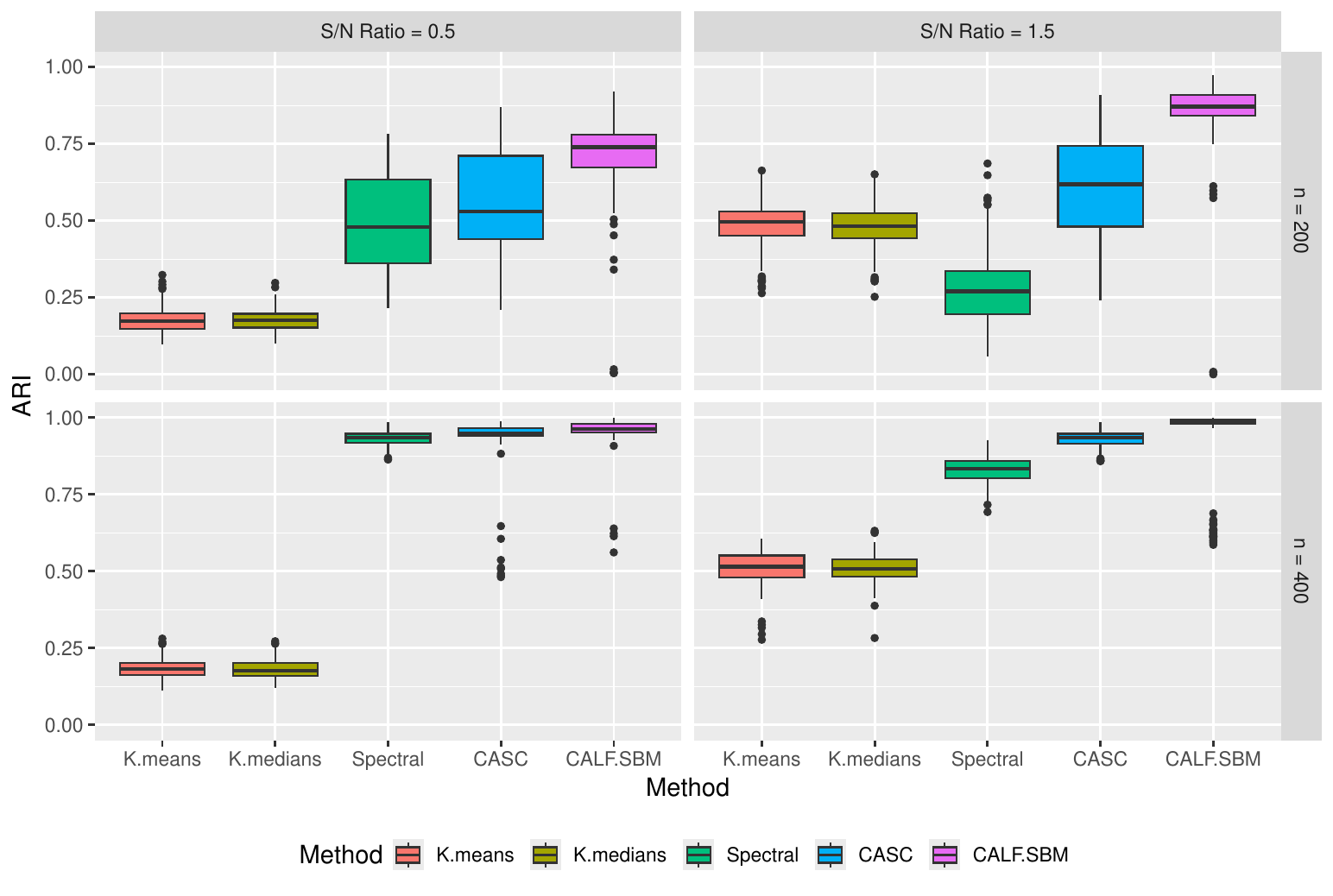}
			\caption{Comparisons of clustering results between the 
			proposed
				CALF-SBM and competing methods using ARI with network 
				size
				$n \in \{200, 400\}$ and signal-to-noise ratios (S/N 
				ratio)
				$\omega \in \{0.5, 1.5\}$.}
			\label{fig:boxplot_knownK}
		\end{figure}

		Further, we compare the proposed CALF-SBM with a collection of
		competing methods that require the knowledge of the true 
		value of
		$K$ for community detection. Specifically, we consider 
		$k$-means
		algorithm, $k$-medians algorithm, spectral clustering and
		covariate-assisted spectral
		clustering~\citep[CASC,][]{binkiewicz2017covariate}. We
		evaluate the performance of each method through the accuracy 
		of node
		membership assignment, quantitatively measured by the
		Adjusted Rand Index~\citep[ARI,][]{rand1971objective}, which 
		is
		ranged between $0$ and $1$ with $0$ and $1$ indicating random
		assignment
		and perfect matching, respectively. For our simulation,
		ARI is particularly preferred since it is invariant to
		label-switching and also robust against unbalanced cluster 
		sizes.
		Figure~\ref{fig:boxplot_knownK} exhibits side-by-side box 
		plots for
		the considered models with varying network size $n \in \{200, 
		400\}$
		and varying signal-to-noise ratio levels (quantified through
		$\omega \in \{0.5, 1.5\}$) for additional sensitivity 
		investigation.

		The results in Figure~\ref{fig:boxplot_knownK} suggest that 
		the
		overall performance of the proposed CALF-SBM is better than 
		the
		competing methods regardless of network size or 
		signal-to-noise
		ratio level. \red{Specifically, when network size is small and
			signal-to-noise ratio is weak, CALF-SBM outperforms the 
			competing
			methods reflected in more accurate clustering as well as 
			smaller 
			variations, although, at the same time, CALF-SBM 
			undergoes with a 
			drawback of more outliers. CACS is slightly more accurate 
			than 
			spectral clustering, whereas $k$-means and $k$-medians
			algorithms do not perform well since they fail to account 
			for
			observed network structure throughout clustering. When 
			network size
			is small but cluster signal becomes stronger, the 
			accuracy of 
			CALF-SBM is notably improved and the variance is 
			significantly 
			reduced, while the improvement in the accuracy of CASC is 
			limited. 
			All the other three methods perform similarly in spite of 
			great 
			improvements in $k$-means and $k$-medians. When network 
			size 
			increases, the proposed CALF-SBM, in average, provides 
			perfect 
			matching results with minimal variations, but does not 
			seem to be as 
			robust as spectral clustering and CASC due to a greater 
			number of 
			outliers. When the signal-to-noise ratio increases from 
			$0.5$ to 
			$1.5$, the variance of CALF-SBM decreases further, while 
			the number 
			of outliers does not change much. Meanwhile, the accuracy 
			of spectral 
			clustering and CASC decreases slightly. Nonetheless, 
			$k$-means and 
			$k$-medians are less preferred owing to relatively low 
			accuracy.}

		\red{In addition to ARI, we adopted normalized mutual 
			information~\citep[NIM,][]{danon2005comparing} to 
			evaluate the 
			clustering performance of CALF-SBM and compared 
			it with the considered competing methods. Unlike ARI, 
			which measures 
			the agreement between clustering assignments based on 
			pairwise 
			comparisons, NMI uses information theory to quantify the 
			mutual 
			dependence between clustering assignments. NMI scores 
			range from $0$ 
			to $1$ with $0$ and $1$ respectively indicating no 
			similarity and 
			perfect similarity. The use of NMI can provide a more 
			comprehensive 
			assessment of CALF-SBM. We provide NMI-based side-by-side 
			box plots 
			to compare CALF-SBM with the four competing methods in 
			Figure~\ref{fig:boxplot_knownK_nmi} in~\ref{app:knownK}, 
			and 
			find that CALF-SBM outperform all competing methods in a 
			similar 
			manner as discussed for Figure~\ref{fig:boxplot_knownK}.}

		\subsubsection{Unknown $K$}
		\label{sec:unknownK}

		The second simulation study aims to assess the performance of 
		the
		proposed CALF-SBM for unknown $K$, alongside comparisons with 
		the
		standard SBM~\citep{nowicki2001estimation} which ignores 
		node-level
		covariates, and a recently developed Bayesian community 
		detection
		model utilizing covariates~\citep[BCDC,][]{shen2023bayesian}.
		Analogously, all the simulation results presented in this 
		section
		are based on $100$ independent replicates with fixed $n = 
		400$ and
		$\omega = 1.5$ suggesting a relatively strong signal-to-noise 
		ratio
		level. Statistical inferences for both (standard) SBM and 
		BCDC are
		drawn using MCMC algorithms which can be practically 
		implemented via
		the utility functions from the \proglang{R} package
		\pkg{bcdc}~\citep{shen2023bayesian}. The specific MCMC
		configurations for all three models are given in
		Table~\ref{tab:mcmc_config}, where the burn-in and iteration 
		numbers
		for BCDC are based on the default settings in the associated
		functions. As SBM does not account for node-level covariates
		throughout clustering, our experiments show that the burn-in 
		number
		of $500$ is sufficiently large for the algorithm to converge 
		when
		network size is $n = 400$.

		\begin{table}[tbp]
			\centering
			\caption{A summary of MCMC configuration setup for 
			CALF-SBM, SBM
				and BCDC.}
			\label{tab:mcmc_config}
			\setlength{\tabcolsep}{16.9pt}
			\begin{tabular}{lrrr}
				\toprule
				& CALF-SBM & SBM & BCDC \\
				\midrule
				Burn-in & 5000 & 500 & 5000 \\
				Iteration & 10000 & 1500 & 15000 \\
				Thinning & 10 & 1 & 1 \\
				\bottomrule
			\end{tabular}
		\end{table}

		We first evaluate all three methods through their 
		performances of
		the selection of true community number $K$. Specifically, we
		considered two sets of true $K$ values: $K = 3$ and $K = 4$. 
		For
		CALF-SBM and SBM, we selected the optimal $K$ via model 
		selection
		from $K \in \{2, 3, \ldots, 6\}$. Specifically, we adopted 
		WAIC for
		CALF-SBM as elucidated in Section~\ref{sec:estK}
		and the Bayesian information criterion (BIC) for SBM. The 
		selection
		of $K$ for BCDC followed the Bayesian framework established
		by~\citet{liu2024variational}. The dynamic selection of $K$ 
		for BCDC
		requires a pre-specification of the concentration parameter 
		for the
		Chinese Restaurant Process used in their algorithm. This
		concentration parameter in our simulations was determined by 
		the
		method of moments given the true value of $K$. The results are
		summarized in Table~\ref{tab:k_selection}, where we observe 
		that the
		proposed method for CALF-SBM has a higher probability of 
		selecting
		the correct $K$ than the two competing approaches, both of 
		which
		tend to overestimate $K$.

		\begin{table}[tbp]
			\centering
			\caption{A summary of the selection of community number 
			$K$ among
				CALF-SBM, SBM and BCDC for $n = 400, \omega = 1.5$.}
			\label{tab:k_selection}
			\setlength{\tabcolsep}{7.8pt}
			\begin{tabular}{l rrrr rrrr rrrr}
				\toprule
				& \multicolumn{6}{c}{True $K = 3$}
				& \multicolumn{6}{c}{True $K = 4$} \\
				\cmidrule(lr){2-7} \cmidrule(lr){8-13}
				Method & 2 & 3 & 4 & 5 & 6 & 7 &
				2 & 3 & 4 & 5 & 6 & 7 \\
				\midrule
				CALF-SBM & 0 & \blue{48} & 9 & 15 & 28 & -- & 0 & 1 & 
				\blue{60} &
				23 & 16 & -- \\
				SBM & 0 & 0 & 17 & 38 & \blue{45} & -- & 0 & 1 & 13 & 
				29 &
				\blue{57} & -- \\
				BCDC & 0 & 0 & 12 & \blue{51} & 17 & 0 & 0 & 0 & 11 & 
				39 &
				\blue{40} & 10 \\
				\bottomrule
			\end{tabular}
		\end{table}

		\begin{figure}[tbp]
			\centering
			\includegraphics[width=0.9\textwidth]{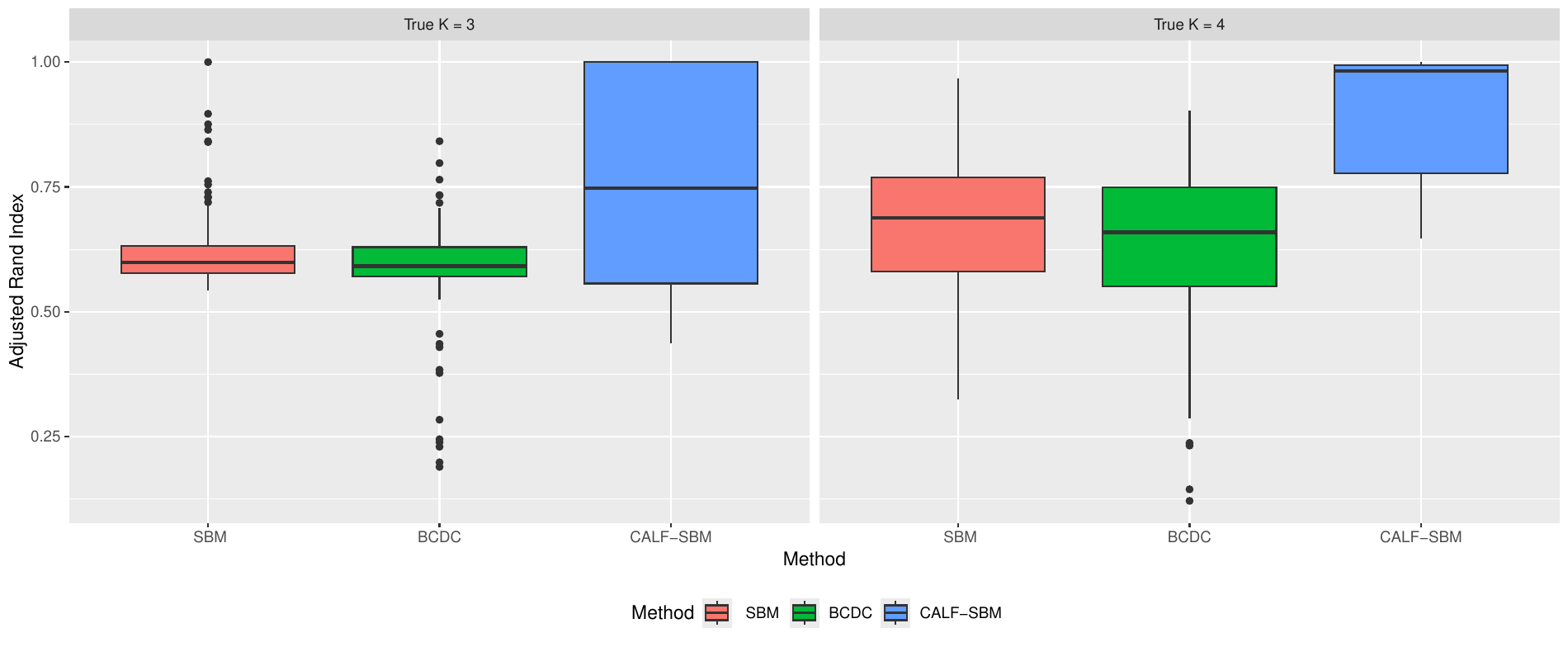}
			\caption{Comparisons of clustering results between the 
			proposed
				CALF-SBM and competing methods using ARI with $n = 
				400$,
				$\omega = 1.5$ and varying $K \in \{3, 4\}$.
			}
			\label{fig:boxplot_unknownK}
		\end{figure}

		Next, we assess the method performances for community 
		detection
		using box plots analogous to the strategy demonstrated in
		Section~\ref{sec:knownK}. The results are shown in
		Figure~\ref{fig:boxplot_unknownK}. For $K = 3$, CALF-SBM 
		outperforms
		SBM and BCDC on average, but undergoes a much greater 
		variance. SBM
		and BCDC present comparative performances, but SBM appears 
		slightly
		preferable to BCDC thanks to fewer ARI outliers on the lower 
		end.
		For $K = 4$, CALF-SBM is, once again, better than SBM and 
		BCDC on
		average. In fact, the median ARI for CALF-SBM is close to $1$ 
		which
		refers to a perfect matching (with respect to the ground 
		truth). The
		improvement of the ARI values (from $K = 3$ to $K = 4$)
		is related to a higher frequency of selecting the correct $K$ 
		as
		shown in Table~\ref{tab:k_selection}. Moreover, we see a 
		substantial
		reduction in the variance of ARI for CALF-SBM with $K = 4$ 
		(compared
		to $K = 3$), where the variances of ARI for both SBM and BCDC
		increase. This is most likely caused by the smaller cluster 
		sizes,
		leading to higher fluctuation in the fit. \red{At last, 
			Figure~\ref{fig:boxplot_unknownK_nmi} 
			in~\ref{app:unknownK} presents 
			the NMI-based side-by-side box plots corresponding to 
			Figure~\ref{fig:boxplot_unknownK} as shown above.}

		\section{Real Data Applications}
		\label{sec:app}
		
		In this section, we demonstrate the utility of CALF-SBM 
		through two
		real data applications. The first application explores a 
		collaboration
		network of otolaryngologists in the United States, utilizing 
		data
		recently collected by researchers at Vanderbilt University 
		Medical
		Center. The second application examines an aviation network, 
		which 
		was
		initially studied to assess the accessibility, in terms of 
		airline
		travel time, of cities in Canada and the United
		States~\citep{frey2007affinity}.

		\subsection{Otolaryngologist Collaboration Network}
		
		We applied the CALF-SBM to a network of otolaryngologist 
		collaboration
		(OCN), with data collected and maintained by a subset of the 
		authors
		at Vanderbilt University Medical Center. The network 
		encompasses all
		$129$ U.S. residency programs accredited by the Accreditation 
		Council
		for Graduate Medical Education. Comprehensive data on $2494$
		physician faculty members, including names, genders, 
		institutional
		affiliations, and specialties, were compiled through 
		extensive web
		scraping. Our Python scripts facilitated the identification 
		of each
		physician's Scopus ID, with subsequent human review for
		disambiguation. We acquired publication data for those with 
		valid
		Scopus IDs via multiple Elsevier APIs, collecting all 
		Scopus-indexed
		articles authored by the identified physicians up to December
		2023. The dataset, derived from publications co-authored by
		at least two identified physicians, included $2259$ 
		physicians and
		$88245$ collaborative relationships.

\begin{figure}[tbp]
			\centering
			\includegraphics[width = 0.9\textwidth]{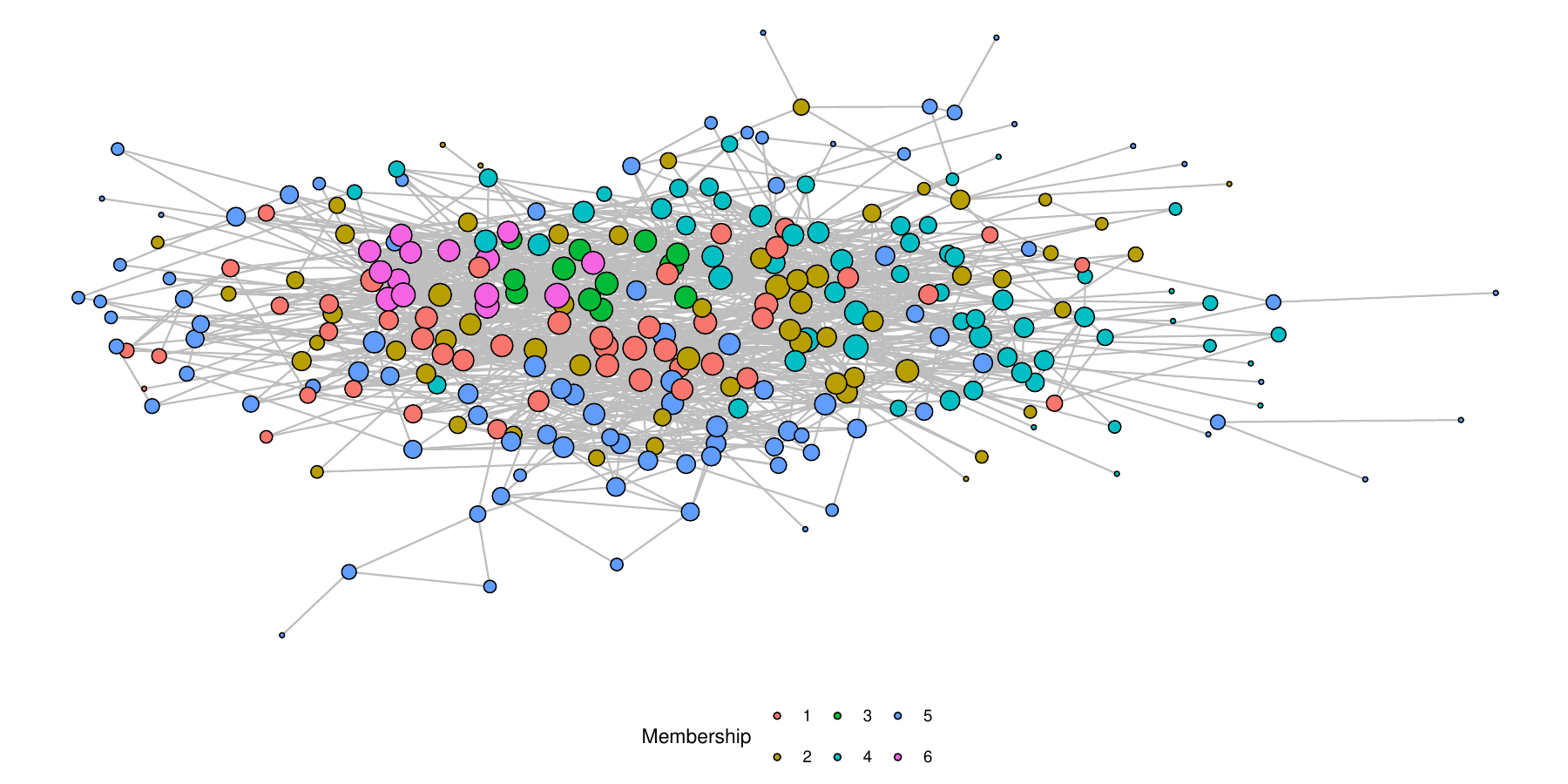}
			\caption{Clustering results for OCN based on the 
			application of
				CALF-SBM ($K = 6$), where the node sizes are 
				proportional to 
				the	logarithm of their degrees.}
			\label{fig:otonet_total}
		\end{figure}

		The CALF-SBM application utilized two covariates for each 
		physician:
		institutional affiliation and specialty. The specialties were
		categorized into seven types: Laryngology, Rhinology, Head \& 
		Neck,
		Neurotology, Facial Plastics, General, and Pediatric. To 
		enhance the
		interpretability of results and maintain computational 
		feasibility, 
		we
		filtered the network data to include only physicians from 
		institutions
		with at least 20 occurrences. This resulted in a final dataset
		comprising $309$ physicians affiliated with $12$ medical 
		schools or
		hospitals, and $2630$ collaborative relationships. For each 
		physician
		pair $i$ and $j$, we computed $S_{ij}$, the similarity 
		measure, based
		on the average match of their institutions and specialties, 
		formulated as $[\bm{1}(\text{institute}_i = 
		\text{institute}_j) + 
		\bm{1}(\text{specialty}_i = \text{specialty}_j)]/2$, where 
		$\bm{1}(\cdot)$ is the standard indicator function. We 
		configured 
		the burn-in and iteration numbers at $10000$ and
		$30000$, respectively, to ensure the convergence of the MCMC
		algorithm, verified by the trace plots of the posterior 
		samples. We
		evaluated potential cluster numbers ranging from $K = 2$ to 
		$K = 7$,
		ultimately selecting $K = 6$ as optimal based on the WAIC. The
		clustering results (generated using~\citet{pkg:tidygraph}) 
		are shown 
		in Figure~\ref{fig:otonet_total}, where the node size is 
		proportional to the logarithm of node degree. Cluster~$5$ is 
		the 
		largest, predominantly comprising physicians with fewer 
		connections 
		within the network. Conversely, Cluster $3$, while being the 
		smallest, includes physicians who are densely connected, 
		indicating 
		strong collaborative ties.

		\begin{table}[tbp]
			\centering
			\caption{Summary of community properties by applying 
			CALF-SBM,
				including community size, dominant institution, 
				dominant
				specialty, community density and maximum (node) 
				degree.}
			\label{tab:otonet}
			\footnotesize
			\setlength{\tabcolsep}{7.0pt}
			\begin{tabular}{l rrrrr}
				\toprule
				& Size & Institution & Specialty & Density & Max Deg. 
				\\
				\midrule
				Cluster 1 & 46 & Michigan ($35\%$) & Head \& Neck 
				($30\%$) &
				0.14 & 22 \\
				Cluster 2 & 59 & Harvard ($19\%$) & Head \& Neck 
				($22\%$) &
				0.08 & 13 \\
				Cluster 3 & 12 & Northwestern ($33\%$) & Rhinology 
				($50\%$)
				& 1.00 & 11 \\
				Cluster 4 & 63 & Mount Sinai ($33\%$) &  General 
				($33\%$) &
				0.09 & 18 \\
				Cluster 5 & 115 & Mount Sinai ($24\%$) &  General 
				($18\%$) &
				0.02 & 15 \\
				Cluster 6 & 14 & Michigan ($29\%$) & Pediatric 
				($86\%$) &
				1.00 & 13 \\
				\bottomrule
			\end{tabular}
		\end{table}

		Given the limited information in 
		Figure~\ref{fig:otonet_total},
		further in-depth analysis on each community is presented in
		Table~\ref{tab:otonet}, which specifically contains cluster 
		size, 
		the most frequent institution (in each cluster), the most 
		frequent 
		specialty (in each cluster), cluster density, and the largest 
		node 
		degree (in each cluster). The smallest communities, 
		Clusters~$3$
		and~$5$, exhibit extremely high connectivity among their
		physicians. Specifically, Cluster~$3$ predominantly consists 
		of
		specialists in rhinology (50\%), while Cluster~$5$ is mainly 
		composed
		of pediatric specialists (86\%). Clusters $1$, $2$, and $4$ 
		have
		similar sizes and within-cluster~densities. Cluster~$1$ is 
		primarily
		made up of physicians from the University of Michigan Health 
		System,
		specializing in Head \& Neck and Pediatric, whereas 
		Cluster~$2$
		includes key participants from Massachusetts Eye and Ear
		Infirmary/Harvard Medical School and University of North 
		Carolina
		Hospitals with similar specialties. Montefiore Medical 
		Center/Albert
		Einstein College of Medicine is notably present in both 
		Clusters $1$
		and $2$, with physician memberships significantly influenced 
		by the
		community affiliations of their connections. Cluster~$4$ 
		chiefly
		involves physicians from Icahn School of Medicine at Mount 
		Sinai/New
		York Eye and Ear Infirmary at Mount Sinai, specializing in
		``general.'' Despite being the largest, Cluster~$5$ displays 
		much 
		lower
		density as nearly half of its physicians have $2$ connections 
		or
		fewer, a contrast to other clusters.

		\red{Further, the estimate of $\beta_0$ is $-0.70$, 
		indicating an 
			average probability of connection of $0.33$. However, the 
			actual 
			density of OCN is just $0.03$, which explains that most 
			estimates of 
			$(\beta_{11}, \beta_{12}, \ldots, \beta_{66})$ are 
			negative. As 
			expected, the estimates for intra-cluster coefficients of 
			$\bm{\beta}$ (average $3.19$) are greater than those for 
			inter-cluster coefficients (average $-3.51$). In 
			particular, there 
			are only two positive estimates for $\bm{\beta}$, 
			$\hat{\beta}_{55} = 
			11.37$ and $\hat{\beta}_{66} = 15.99$, which suggests 
			that the 
			connectedness between physicians within Clusters $5$ 
			and~$6$ depends 
			largely on their commonality in institutions and 
			specialties. The 
			estimates of subject-level latent factors, $\bm{\theta}$, 
			are 
			centered around $-0.01$, while the estimate of standard 
			deviation is 
			$1.51$, suggesting that the effects of individual latent 
			factors on 
			connectivity are lower than those of clustering 
			membership. Moreover, 
			we conduct post-hoc analyses to evaluate the Cram\'{e}r's 
			V between 
			clustering membership and institution ($0.36$, 
			Chi-squared test 
			$\text{p} = 0.001$) and the Cram\'{e}r's V between 
			clustering 
			membership and specialty ($0.26$, Chi-squared test 
			$\text{p} = 
			0.001$), respectively. Lastly, by fitting a generalized 
			linear model 
			relating clustering membership and $\bm{\theta}$ 
			estimates, adjusted 
			for institutions and specialties, we find $\text{p} = 
			0.005$, which 
			suggests the need to account for subject heterogeneity in 
			the model.}

		\subsection{Airport Reachability Network}
		
		The second real data application examines an airport 
		reachability
		network (ARN), as outlined by 
		\citep[ARN,][]{frey2007affinity}, which
		includes data on direct flights among major airports in the
		U.S. (covering all continental states and Hawaii) and Canada. 
		The ARN
		comprises a total of $456$ airports. The two covariates 
		utilized are
		the geographical locations of the airports (expressed through 
		latitude
		and longitude) and the log-transformed metropolitan 
		populations of 
		the
		cities hosting these airports. In constructing the similarity 
		matrix,
		two components were considered: the great-circle distances 
		between
		airports, which account for Earth's curvature, and the 
		population
		differences between the associated metropolitan areas. As 
		recommended
		by \citet{shen2023bayesian}, these difference measures were 
		scaled to
		unit variance before being combined with equal weighting to 
		complete
		the similarity matrix.

\begin{figure}[tbp]
			\centering
			\includegraphics[width = 
			0.9\textwidth]{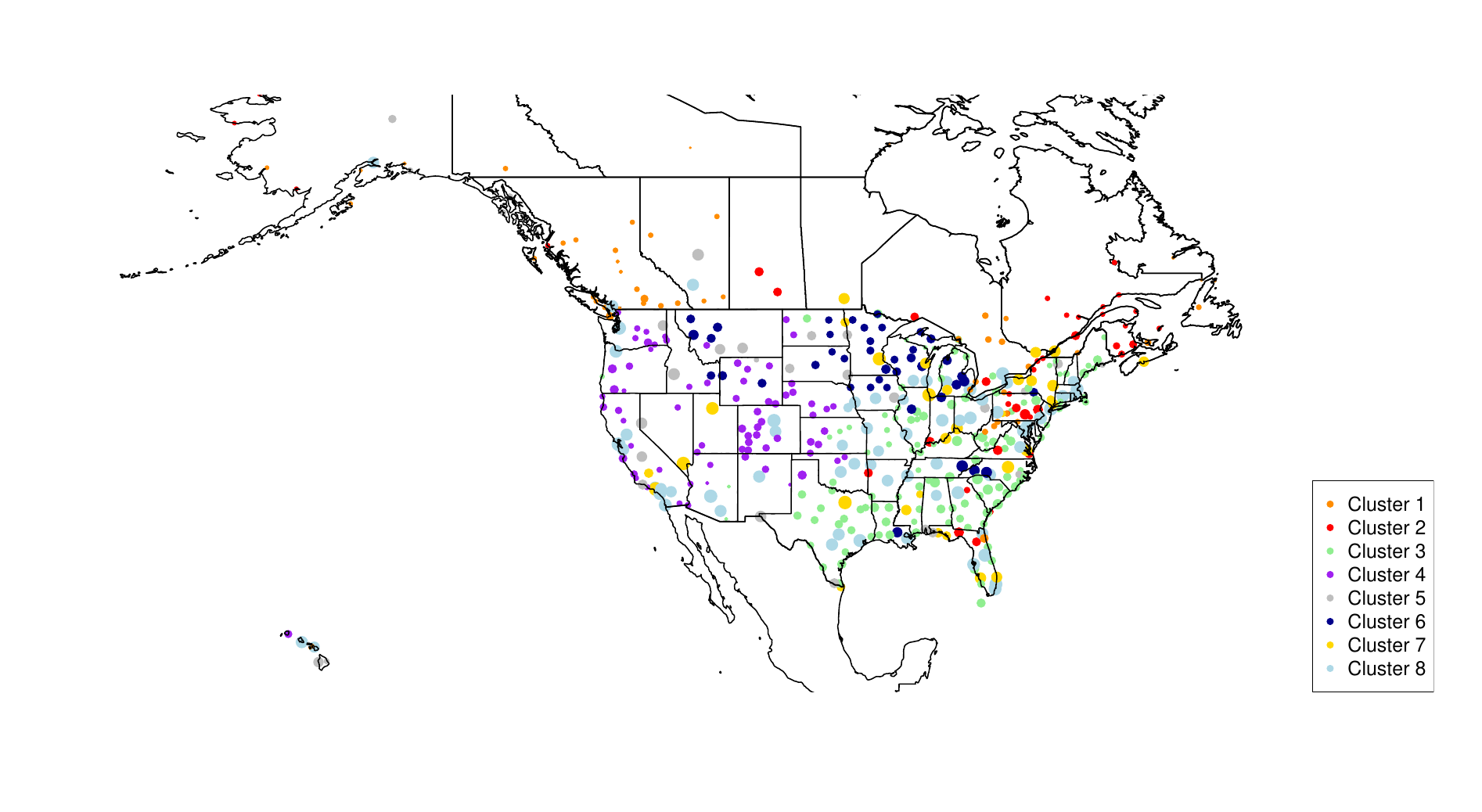}
			\caption{Clustering result based on the application of 
			CALF-SBM 
				to
				the undirected ARN.}
			\label{fig:map}
		\end{figure}

		The original ARN described in \citet{frey2007affinity} is 
		directed. 
		We
		adapted it to an undirected network by ensuring that a link 
		exists
		between two airports if there is at least one direct flight 
		in each
		direction. This modification resulted in an undirected 
		network nearly
		structurally equivalent to the original directed ARN. 
		Typically, if 
		an
		airport \(A\) has direct flights to another airport \(B\), 
		\(B\) 
		often
		has return flights to \(A\). The nature of ARN necessitates 
		the
		inclusion of latent factors (i.e., \(\theta\)) to address the
		significant degree heterogeneity observed, with major hubs 
		connecting
		to hundreds of airports, whereas most airports in smaller 
		cities have
		fewer connections. This degree heterogeneity is illustrated in
		Figure~\ref{fig:map}, where each node's size is proportional 
		to its
		degree.

		To apply CALF-SBM to the ARN, we set the burn-in and 
		iteration 
		numbers
		to $5000$ and $10000$, respectively, ensuring the convergence 
		of the
		MCMC algorithm. We determined that $K = 8$ was the optimal 
		number of
		clusters based on WAIC. The clustering results are also 
		depicted in
		Figure~\ref{fig:map}. Due to the high density of the network, 
		edges
		are omitted in the figure for clarity. Additional information 
		about
		the cluster summaries is provided in Table~\ref{tab:airport}.

		\begin{table}[tbp]
			\centering
			\caption{Summary of community properties of airports 
			across the
				U.S.\ and Canada, including community size, 
				(approximate)
				center, largest airport (in terms of number of 
				connections),
				metropolitan size median and community density.}
			\label{tab:airport}
			\footnotesize
			\setlength{\tabcolsep}{7.0pt}
			\begin{tabular}{l rrrrr}
				\toprule
				& Size & Center & Largest Airport & Median Metro & 
				Density \\
				\midrule
				Cluster 1 & 51 & Saskatchewan & St.\ Johns, NL & 
				46,850 &
				0.18 \\
				Cluster 2 & 41 & Michigan & Harrisburg, PA & 81,445 &
				0.24 \\
				Cluster 3 & 120 & Kentucky & Greenville, SC & 162,052 
				&
				0.26 \\
				Cluster 4 & 72 & Utah & Amarillo, TX & 60,129 &
				0.33 \\
				Cluster 5 & 33 & Idaho & Boise, ID & 186,738 &
				0.55 \\
				Cluster 6 & 41 & Minnesota & Knoxville, TN & 110,138 &
				0.85 \\
				Cluster 7 & 31 & Indiana & Los Angeles, CA & 972,634 &
				0.87 \\
				Cluster 8 & 67 & Kansas & San Francisco, CA & 
				1,795,000 &
				0.97 \\
				\bottomrule
			\end{tabular}
		\end{table}
		
		Table~\ref{tab:airport} suggest that the estimated clusters 
		correlate
		with both the geographical locations of airports and the 
		sizes of the
		cities in which they are situated. Cluster~$1$ predominantly 
		includes
		small airports in Canada, with more than $75\%$ located in 
		cities
		having a metropolitan population under $50,000$. Cluster~$2$ 
		consists
		of small airports in the Eastern U.S. and smaller Canadian 
		airports
		not included in Cluster~$1$. In contrast, Cluster~$4$ 
		comprises small
		airports in the Western U.S., with no Canadian airports. 
		Cluster~$3$
		diverges by encompassing medium-sized airports in the Eastern 
		U.S.\@
		and is the largest of the identified communities. 
		Clusters~$5$ 
		and~$6$
		mainly contain medium to large airports, with Cluster~$5$ 
		being more
		western (centered in Idaho) and Cluster~$6$ more northern 
		(centered 
		in
		Minnesota).

		Clusters~$7$ and~$8$ primarily consist of the largest air 
		hubs in
		North America, exhibiting high internal connectivity. Despite 
		this,
		there is no clear geographical distinction separating these
		clusters. Cluster~$7$, smaller in size, includes fewer hubs 
		located
		predominantly in the central and western U.S.: Chicago, 
		Illinois;
		Dallas/Fort Worth, Texas; Las Vegas, Nevada; Los Angeles, 
		California;
		Minneapolis/St. Paul, Minnesota; and Salt Lake City, Utah. 
		Conversely,
		Cluster~$8$, which is twice the size of Cluster~$7$, contains 
		hubs
		that are more dispersed across the U.S., with a majority on 
		the east
		coast: Atlanta, Georgia; Boston, Massachusetts; New York 
		City, New
		York; Washington, D.C.; the southern U.S.: Denver, Colorado; 
		Phoenix,
		Arizona; and notable west coast hubs: San Francisco, 
		California; and
		Seattle/Tacoma, Washington. Despite San Francisco being the 
		most
		connected with $442$ connections, as noted in 
		Table~\ref{tab:airport},
		Cluster~$8$ also features airports of comparable size, such as
		Washington D.C. (with $432$ connections) and New York City 
		(with 
		$428$
		connections).

		\section{Discussions}
		\label{sec:dis}

		This paper introduces an extended SBM, called CALF-SBM, that
		simultaneously incorporates node-level covariates and nodal
		heterogeneity to improve network clustering. The associated 
		MCMC
		algorithm is capable of generating unbiased estimates for 
		model
		parameters. The implementation of the proposed algorithm is 
		based on
		the \pkg{NIMBLE} package which compiles it using 
		\proglang{C++} for
		speed. The number of communities is assumed unknown, and can 
		be
		accurately estimated using a reliable model selection 
		approach.
		According to our simulation studies, the proposed CALF-SBM
		outperforms popular competing methods for network clustering.
		Applications to the two real network data give new insights 
		into
		their respective community structures. User-friendly 
		functions for
		directly implementing CALF-SBM are available in an 
		\proglang{R}
		package on GitHub.

		\red{In the current project, there is an issue that requires 
		further 
			study. Variable selection regarding node covariates has 
			not been 
			considered in the proposed CALF-SBM, but there may be 
			noise 
			covariates that do not contribute to network generation 
			and 
			potentially affect inference accuracy. Some preliminary 
			experiments 
			show that the presence of noise covariates leads to 
			biased 
			estimates, but the clustering performance is still 
			satisfactory. 
			Nevertheless, systematic research is still needed to 
			address model 
			robustness.} Besides, there are several potential 
			extensions to the 
		current CALF-SBM. First, while the proposed MCMC algorithm is 
		implemented in \proglang{C++}, it remains computationally 
		demanding 
		and faces convergence challenges for large networks. This
		indicates a need for improvements to enhance scalability. 
		\red{One 
			potential consideration is the variational Bayesian 
			approach~\citep{blei2017variational}, but its feasibility 
			and 
			performance (under various scenarios) are not 
			straightforward, 
			deserving a comprehensive study.} Secondly,
		the model is readily extendable to directed networks; 
		however, this
		generalization, though conceptually simple, requires more 
		network 
		data
		for reliable inference, emphasizing further the need to 
		improve
		algorithm scalability for handling big network data. 
		Ultimately, the
		model should also adapt to both weighted and directed 
		networks with
		more accurate inference and richer result
		interpretation. Lastly, by incorporating edge-level attributes
		alongside node-level covariates, the model could become more 
		powerful
		and flexible, potentially yielding more insightful network 
		clustering
		outcomes by leveraging all available data.
		
		\section*{Supplementary Materials}
		
		The online supplement collects the \proglang{R} codes for 
		reproducible simulations in Section~\ref{sec:sim}.

		\bibliographystyle{elsarticle-num-names}

		\appendix
		
		\section{Side-by-Side Box Plots Using NMI for the Simulations 
		in 
			Section~\ref{sec:knownK}}
		\label{app:knownK}
		
\begin{figure}[tbp]
			\centering
			\includegraphics[width=0.9\textwidth]{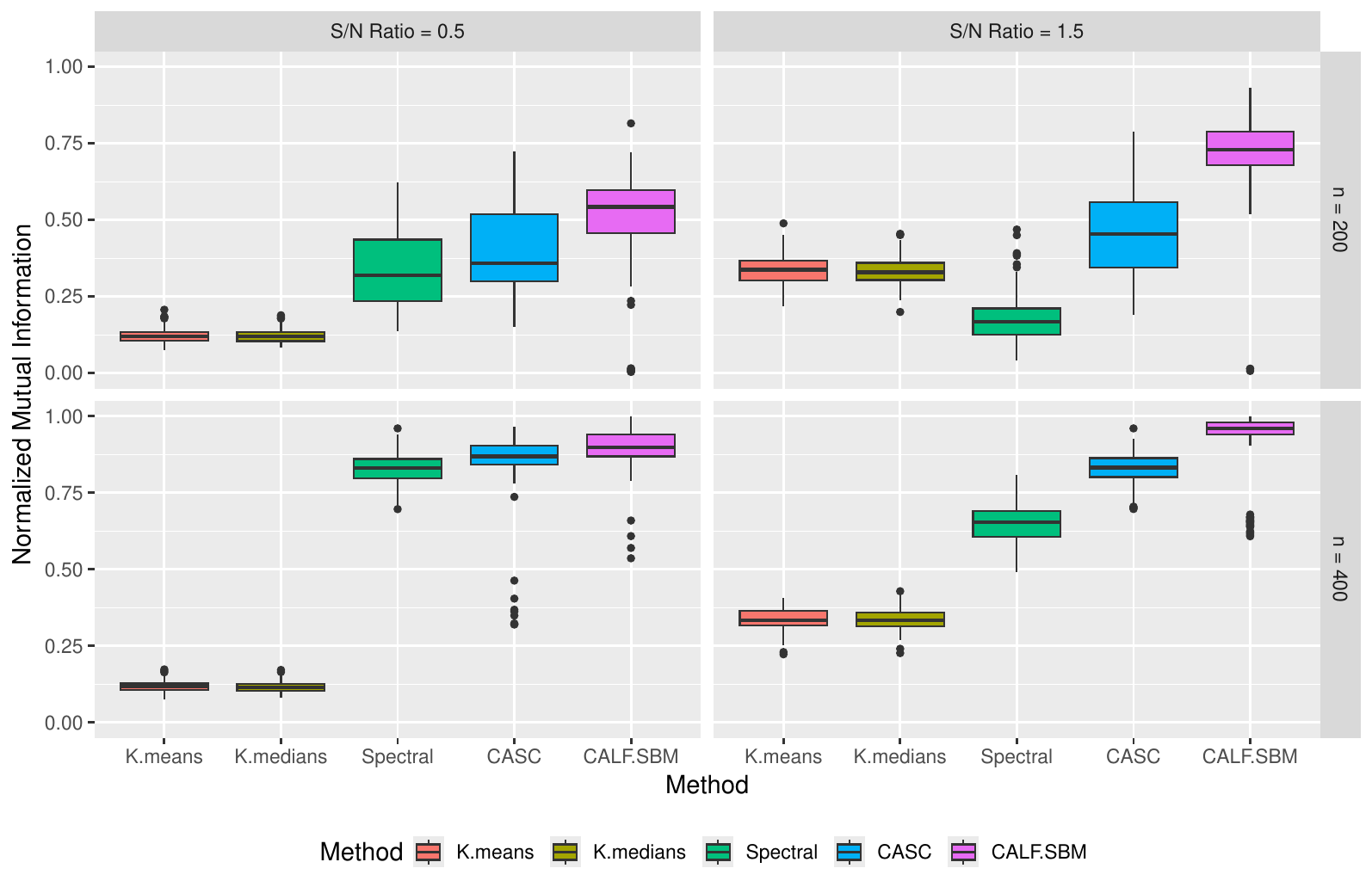}
			\caption{Comparisons of clustering results between the 
			proposed
				CALF-SBM and competing methods using NMI with network 
				size
				$n \in \{200, 400\}$ and signal-to-noise ratios (S/N 
				ratio)
				$\omega \in \{0.5, 1.5\}$.}
			\label{fig:boxplot_knownK_nmi}
		\end{figure}

		\red{Figure~\ref{fig:boxplot_knownK_nmi} shows that the 
		proposed 
			CALF-SBM 
			outperforms the considered competing methods in all 
			scenarios. The 
			overall clustering accuracy of CALF-SBM improves with the 
			increase 
			of network size and signal-to-noise ratio, while the 
			variance of NMI 
			decreases notably. For CALF-SBM, outliers in NMI are 
			still observed, 
			but become less severe when the network size is larger 
			and the 
			signal-to-noise ratio is higher.}
		
		\section{Side-by-Side Box Plots Using NMI for the Simulations 
		in 
			Section~\ref{sec:unknownK}}
		\label{app:unknownK}
		
\begin{figure}[tbp]
			\centering
			\includegraphics[width=0.9\textwidth]{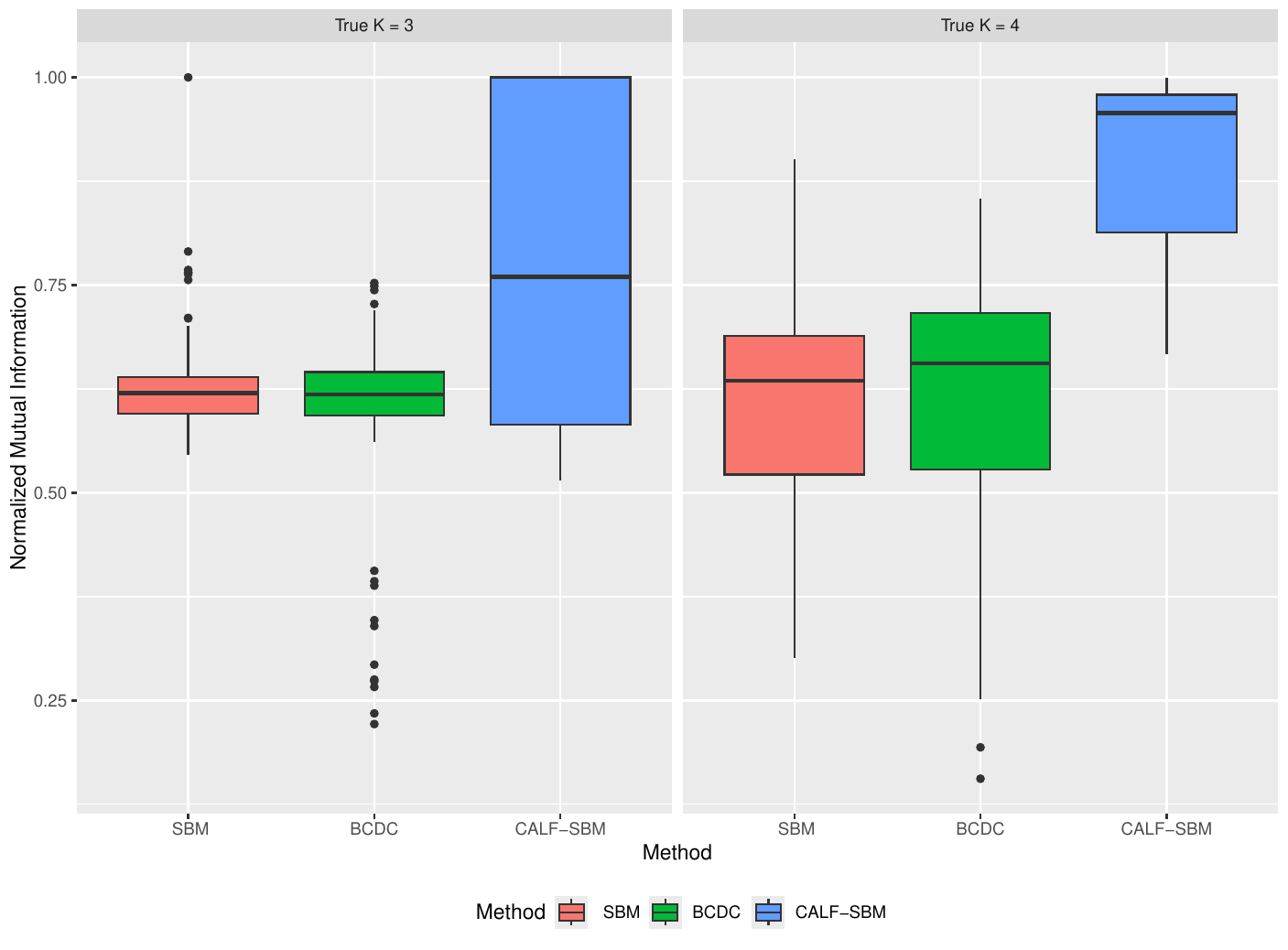}
			\caption{Comparisons of clustering results between the 
			proposed
				CALF-SBM and competing methods using NMI with $n = 
				400$,
				$\omega = 1.5$ and varying $K \in \{3, 4\}$..}
			\label{fig:boxplot_unknownK_nmi}
		\end{figure}

		\red{Figure~\ref{fig:boxplot_unknownK_nmi} shows that the 
		proposed 
			CALF-SBM outperforms the two competing methods in terms 
			of clustering 
			accuracy. Although CALF-SBM experiences a larger NMI 
			variance when $K 
			= 3$, it has fewer outliers than the other two methods. 
			The NMI 
			variance of CALF-SBM decreases with the increase of $K$, 
			while the 
			NMI variances of SBM and BCDC both get larger.}
		
	\end{document}